\documentclass[prl,twocolumn,nofootinbib]{revtex4-1}
\usepackage{amsmath,amssymb,amsfonts}
\usepackage{graphicx}
\usepackage[english]{babel}

\usepackage{bm}
\usepackage[utf8]{inputenc}

\usepackage{listings}

\usepackage[T1]{fontenc}
\usepackage{lmodern}

\usepackage{mathrsfs}
\usepackage{enumerate}

\usepackage[normalem]{ulem}
\usepackage{color}
\usepackage{array}
\def\beq{\begin{equation}}
\def\eeq{\end{equation}}
\def\be{\begin{equation}}
\def\ee{\end{equation}}
\def\bea{\begin{eqnarray}}
\def\eea{\end{eqnarray}}

\begin{document}

\title{Sterile neutrino searches at tagged kaon beams}    
          
\date{\today}
\author{Luis A. Delgadillo}
\email{luisd@vt.edu}
\author{Patrick Huber}
\email{pahuber@vt.edu}
\affiliation{Center for Neutrino Physics, Department of Physics, Virginia Tech, Blacksburg, Virginia 24061, USA}

\begin{abstract}
   Tagged kaon beams are attractive neutrino sources, which would
   provide flavor pure $\nu_e$-beams with exactly measured
   normalization. We point out that this also leads to an anti-tagged
   flavor pure $\nu_\mu$-beam, with equally well known normalization. Exposing a 1\,kt liquid argon detector
   at a baseline of 1\,km to this combination of unique beams allows
   to decisively test recent indications by IceCube and Neutrino-4 of
   sterile neutrino oscillations in the multi-eV range.
\end{abstract}

\maketitle




\section{Introduction}
\label{sec:intro}


The decisive measurement of a nonzero reactor mixing
angle~\cite{An:2012eh,Ahn:2012nd,Abe:2014bwa} has marked the beginning
of an era of precision neutrino mixing measurements. The current neutrino
oscillation data have determined the neutrino mass squared
differences, $\Delta m_{21}^{2}$ and $|\Delta m_{31}^{2}|$, and the
mixing angles $\sin^2(\theta_{ij})$ (with $ij=\{12,13,23\}$) of the
lepton mixing matrix \cite{Zyla:2020zbs}. Nevertheless, there exist
persistent hints from different experiments, notably
LSND~\cite{Aguilar:2001ty} and
MiniBooNE~\cite{Aguilar-Arevalo:2018gpe}, as well as reactor
experiments~\cite{Mention:2011rk,Ko:2016owz,Almazan:2018wln,Alekseev:2018efk,Berryman:2020agd}
that may indicate the existence of a fourth neutrino species, a
sterile neutrino $\nu_{s}$ with a mass of $\mathcal{O}$(1) eV mixing
with active neutrinos. For a review on global sterile neutrino
oscillations see {\it e.g.}~\cite{Dentler:2018sju} and references
therein. While sterile neutrino oscillation would be a simple
explanation, if applied to both the $\nu_\mu$-appearance results
(LSND, MiniBooNE) and the reactor indications, it also would predict
the disappearance of $\nu_\mu$, which has not been observed. On the
contrary, strong limits have been placed on this mode over the years
and thus a sterile neutrino interpretation of all these anomalous
results seems unlikely. For masses of the new state above an eV, also
cosmological bounds become an issue and direct bounds from KATRIN
apply as well~\cite{Aker:2019uuj}. More recently two experiments
provide indications of multi-eV scale oscillations. Specifically, the
Neutrino-4 experiment reports a best-fit $\Delta m^2\simeq
7\,\mathrm{eV}^2$~\cite{Serebrov:2020kmd} and IceCube reports a best
fit value of $\Delta m^2\simeq
4.5\,\mathrm{eV}^2$~\cite{PRLIce,Aartsen:2020iky}.  The multi-eV region is
difficult to access for reactor neutrino experiments due to the
smearing out of oscillations because of the reactor core size. It is not
yet clear if these new indications are compatible with the other
eV-scale indications and if they will persist with increased
statistics and/or a more careful assessment of systematic
uncertainties. In this paper we will investigate tagged kaon beams~\cite{Bernstein:1990bk},  to
study multi-eV scale neutrino oscillation directly and with extremely
well constrained systematics.

A recent example of a proposal to build a tagged kaon beam is the
ENUBET (Enhanced NeUtrino BEams from kaon Tagging) beamline technology~\cite{Longhin:2014yta} based on the reconstruction of positrons from the three-body
semi-leptonic $K^{+}\rightarrow e^{+} \nu_{e} \pi^{0}$ decay, aimed to
determine the absolute $\nu_e / \nu_{\mu}$ flux at $1\%$ level. The
positrons are identified in the decay tunnel by calorimetric
techniques and the beam-line is optimized to enhance the $\nu_{e}$-component
from the three-body semi-leptonic decay and suppress to a negligible
level the $\nu_{e}$-contamination from muon decays
\cite{Longhin:2014yta}. In ENUBET, the rate of positrons provides a
direct measurement of the $\nu_{e}$ produced in the tunnel.

Neutrinos from this type of source will oscillate as usual, since the precision of neither the time nor the energy measurement on the positron will allow to determine the mass eigenstate produced and thus all that tagging does is to fix the beam normalization and baseline travelled.


The Neutrino-4 indication is observed in the $\bar\nu_e\rightarrow\bar\nu_e$ channel and the IceCube indication is observed in the $\nu_\mu\rightarrow\nu_\mu$  channel only, thus no specific prediction for an effect in $\nu_\mu\leftrightarrow\nu_e$ channels arises. A direct test of those indications is therefore best obtained by using the \emph{same} channels as the original results\footnote{Note, that as long as CPT symmetry holds  $\bar\nu_\alpha\rightarrow\bar\nu_\alpha$ has to have the same oscillation probability as $\nu_\alpha\rightarrow\nu_\alpha$.}. Therefore, we will focus the analysis here on these two disappearance channels.

\section{Experimental Framework}
\label{sec:frame}

In this section we describe the experimental set up and the
assumptions that we use in the present analysis. The predicted event
rates were calculated based on neutrino fluxes provided by the ENUBET
collaboration \cite{Longhin:2014yta, Bru}. We consider a 1kt liquid
argon detector with energy resolution which follows a Gaussian
distribution with a width of $\sigma(E)=17\%/\sqrt{E [\mbox{GeV}]}$
for electrons and $\sigma(E)=10\%/\sqrt{E [\mbox{GeV}]}$ for muons, a
total of 50 bins in the energy interval of $0$-$10$ GeV were
considered \cite{Acerbi:2019qiv,Bru}. All calculations are performed
with GLoBES \cite{Huber:2004ka,Huber:2007ji} using the N-flavor
oscillation engine of Refs.~\cite{Kopp:2006wp,Kopp:2007rz}.

\par The signal is obtained from the survival probability of electron
neutrinos ($\nu_{e} \rightarrow \nu_{e}$) stemming from the $K^{+}$ in
the beam that decay into $e^{+}+\nu_e$, which then interact in the
liquid argon detector through the weak charged current. The yields of
kaons transported to the entrance of the decay tunnel is 1.69$\times
10^{-3}$ $K^+/$PoT for 120 GeV protons. The tagged $\nu_{e}$-flux is
assumed to be $99\%$ pure. The largest source of beam related
background in the detector is due to neutral current coherent $\pi^0$ production: The
$\pi^0$ decays into two photons and for pion energies below $\sim
1$\,GeV the opening angle is large enough to cleanly reconstruct both
photons and thus no confusion with a $\nu_e$ charged current event
arises. At higher energies, however, the two photons are more
collinear and may no longer be reconstructed as two particles, hence
these neutral currents events may be misidentified as charged current
$\nu_{e}$ events. Liquid Argon (LAr) detectors have very fine
granularity and as a result very good particle identification. In
particular photon-induced showers can be recognized by the gap between
the vertex and the start of the shower. Without going into the details
of event reconstruction, we estimate the rate of coherent, neutral
current $\pi^0$ production. The cross section for neutral current
coherent $\pi^0$ production has been measured by MINOS on iron
\cite{Adamson:2016hyz} and by NOvA on carbon \cite{Acero:2019qcr}; we
use a theory-derived scaling factor of $(A/12)^{2/3}$ to translate these
results for argon in accordance with the Berger-Sehgal model
\cite{Berg}. Expected $\pi^0$ rates were found to be of order 0.1$\%$
compared to our signal $\nu_{e}$ events and thus can be neglected for
this analysis. The beam normalization is know at the 1\%-level due to
the high kaon tagging efficiency~\cite{Longhin:2014yta}. 

\par Similarly, muon neutrinos can be selected at the neutrino
detector using radius-energy correlations. We performed a 5 GeV energy
cut to avoid contributions from un-tagged $\pi^+$; since the the
branching ratio for the semi-leptonic decay $K^{+} \rightarrow e^+
\nu_{e} \pi^0$ is well known, 5$\%$, and the number of kaon decays in
this mode is fixed by tagging, we can use the equally well-know
branching ratios for the muon neutrino generating decay modes $K^{+}
\rightarrow \mu^+ \nu_{\mu}$ (60$\%$) and $K^{+} \rightarrow \mu^+
\nu_{\mu} \pi^0$ (3\%) to know the muon neutrino flux with the same
accuracy as the electron neutrino flux. Final states here are charged current $\nu_{\mu}$
interactions with the detector. As in the electron neutrino case, neutral
current background events were found to be of order 0.1$\%$ compared
to signal and thus, negligible.

\begin{figure}[t]
\label{fig:1}
\includegraphics[width=0.46\textwidth]{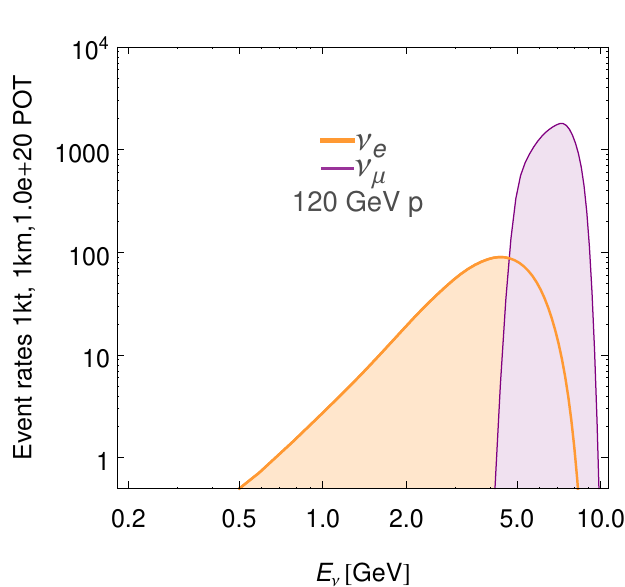}
\centering
\caption{\label{fig1} Expected event rates assuming a 1kt liquid argon detector at baseline $L=1$km for 120 GeV protons with a power beam of $10^{20}$ PoT/yr. }
\end{figure}

Placing a 1\,kt liquid argon detector at a distance of 1\,km from the decay pipe we obtain 1568 $\nu_e$
events and 24603 $\nu_{\mu}$ events/year\footnote{This includes cuts on the beam radius with acceptances of 24\% for $\nu_{e}$ and 34\% for $\nu_\mu$, respectively~\cite{nu2020}.}. Detection efficiencies in this type of detector are close 100\% and for simplicity we neglect them, a simple increase of running time or detector mass by 10-20\% will be required to compensate for this approximation. For a few-GeV beam, a distance of
$L=1$km yields $L/E_{\nu}$ $\simeq$ 0.2 [km/GeV], which corresponds to
an oscillation with $\Delta m^2\simeq
5\,\mathrm{eV}^2$. The $\nu_{\mu}$
energy $E_{\nu}$ spectrum peaks at approximately $7$\,GeV, while the  $\nu_{e}$ energy spectrum peaks around 4\,GeV as shown in
Fig.~[\ref{fig1}]. We do not include the appearance channels in our analysis and have confirmed that their inclusion would not impact our results in an appreciable manner.

\section{Sterile neutrino searches} 
\label{sec:num}

Our treatment of the $3+1$ framework of neutrino oscillations follows
the leptonic mixing matrix parameterization \cite{Dentler:2018sju}
\begin{equation}
\label{par}
U=R(\theta_{34})\tilde{U}(\theta_{24},\delta_{24})R(\theta_{14})R(\theta_{23})\tilde{U}(\theta_{13},\delta)\tilde{U}(\theta_{12},\delta_{12}),
\end{equation}
were $R(\theta_{i,j})$ are orthogonal $4\times4$ matrices on the
($i,j$)-plane, $\tilde{U}(\theta_{i,j},\delta_{i,j})$ are $4\times4$
unitary matrices on the ($i,j$)-plane and $\delta$ is the standard
Dirac CP violating phase, under the short baseline approximation all
extra phases are zero $i.e.$ there will be no additional CP violation in the $3+1$
scenario. The probability for a neutrino produced in the flavor
eigenstate $\nu_{\alpha}$ to be observed as flavor $\nu_{\beta}$ after
traveling some distance $L$ in vacuum and having energy $E$ is:
\begin{equation}
\label{prob}
\begin{split}
    P_{\alpha \beta} & =\delta_{\alpha \beta}-4\sum_{i>j}\mathcal{R}\Big[U^{*}_{\alpha i} U_{\alpha j}U_{\beta i} U^{*}_{\beta j}\Big] \sin^2 \big( 1.27 \Delta m^2_{i j} \frac{L}{ E} \big) \\
   &~~~~~~ +2\sum_{i>j}\mathcal{I}\Big[U^{*}_{\alpha i} U_{\alpha j}U_{\beta i} U^{*}_{\beta j}\Big] \sin \big( 2.54 \Delta m^2_{i j} \frac{L}{E} \big),
\end{split}
\end{equation}
where $U_{\alpha i}~ (\alpha= e, \mu, \tau, s; i= 1,2,3,4) $ are the
elements of the leptonic mixing matrix, $E$ is the neutrino energy,
$L$ is the beam to detector distance, $\Delta m_{i j}^2=m_i^2-m_j^2$
are the squared mass splittings between the standard neutrino mass
eigenstates $\nu_{1}$, $\nu_{2}$, $\nu_{3}$ and a $\nu_4$ sterile
state.  

\begin{table}[h]
\caption{\label{tab:2}Relevant oscillation parameters in the 3+1 scenario used in this analysis.}
\centering
\begin{tabular}{c c c c}
\hline \hline
standard PMNS & value [\textbf{NO}] & sterile parameter &  value \\
\hline 
$\theta_{12}$ & 33.2$^{\circ}$  & $\delta_{24}$ & 0\\
$\theta_{23}$ & 45$^{\circ}$ & $\delta_{12}$ & 0\\
$\theta_{13}$ &  9$^{\circ}$ & $\theta_{34}$ & 0\\
$\Delta m^2_{21}$ [10$^{-5}$eV$^2$] & 7.5  & $\theta_{24}$& free\\ 
$|\Delta m_{31}^2|$ [10$^{-3}$eV$^2$] & 2.6 & $\theta_{14}$& free\\ 
$\delta$ & 0 & $\Delta m^2_{41}$ [10$^{-1}-$10$^2$eV$^2$] & free\\
\hline \hline
\end{tabular}
\end{table} 

Based on our simulated charged current event rates and assuming a
$10^{20}$ PoT/yr beam power on a $1$kt LAr detector, we obtain
sensitivities for the hypothesis of sterile neutrino oscillation under
a (3$+$1) scenario assuming five years of beam operation.
Interpretation of the ENUBET experimental data in terms of sterile
neutrino oscillations allows to test large values of $\Delta m_{41}^2$
and relatively sizable mixing between $\nu_{e}$ and $\nu_{s}$
states. This corresponds well with the parameter space regions
indicated by the gallium results~\cite{Kostensalo:2019vmv}, Neutrino-4
results~\cite{Serebrov:2020kmd} and IceCube
results~\cite{PRLIce,Aartsen:2020iky}. Sensitivity contours were calculated
based on the $\Delta \chi^2$ value for each parameter pair ($\Delta m_{41}^2$, $\sin^2 2
\theta_{14}$) and by determining the boundary of the corresponding
exclusion/allowed regions by translating the $\Delta\chi^2$ to
confidence levels using a $\chi^2$-distribution with two degrees of
freedom. For a recent discussion about the limitations of Wilks' theorem
in disappearance searches with free beam normalization see
Ref.~\cite{Coloma:2020ajw}, which is not quite the same case as
considered here, but gives an indication of the size of the resulting
corrections. We consider different null hypotheses $H_0$:
\begin{itemize}
  \item $H_0$: no oscillation. We compute data assuming no
    disappearance and fit the resulting Asimov data set with finite
    value of $\Delta m^2$ and $\sin^22\theta$. The result is a sensitivity limit, shown as dashed lines.
  \item $H_0$: oscillation according the RAA best fit. We compute data assuming the best fit of the RAA is true
    and fit the resulting Asimov data set with different
    value of $\Delta m^2$ and $\sin^22\theta$. The result are allowed regions (with closed contours), shown as solid lines.
  \item $H_0$: oscillation according the Neutrino-4 best fit. We compute data assuming the best fit of Neutrino-4 is true
    and fit the resulting Asimov data set with different
    value of $\Delta m^2$ and $\sin^22\theta$. The result are allowed regions (with closed contours), shown as solid lines.
      \item $H_0$: oscillation according the IceCube best fit. We compute data assuming the best fit of IceCube is true
    and fit the resulting Asimov data set with different
    value of $\Delta m^2$ and $\sin^22\theta$. The result are allowed regions (with closed contours), shown as solid lines.    
\end{itemize}
For all cases we consider a combined fit of muon and electron neutrino
disappearance and profile over the not-shown $\theta_{i4}$ mixing
angles.

Here, the normalization error of the signal becomes important, while
the beam flux is very well know due to tagging the signal charged
current cross sections are subject to large uncertainties. We assume,
that the main physics program of ENUBET, which are cross section
measurements, have reduced the resulting effective signal uncertainty
to either 1\%, 2\% or 5\%, where we take 2\% as default unless stated
otherwise.


The effective two-flavor limit in the electron disappearance channel yields this simple
oscillation probability:
\begin{equation}
\label{prob1}
P_{e e} = 1-4\vert U_{e4}\vert^2 \big(1- \vert U_{e4}\vert^2\big) \sin^2 \big( 1.27 \Delta m^2_{41} \frac{L}{ E} \big),
\end{equation}
and according to the parameterization Eq.(\ref{par}) this case reduces to the effective two flavor oscillation.
\begin{equation}
\label{prob2}
P_{ee}=1-\sin^2 2 \theta_{ee} \sin^2 \big( 1.27 \Delta m^2_{41} \frac{ L}{ E} \big),
\end{equation}
where $\sin^2 2 \theta_{ee}=4\vert U_{e4}\vert^2 \big(1- \vert
U_{e4}\vert^2\big)$ and $\theta_{ee}=\theta_{14}$ is the angle that
encodes mixing. 
\begin{figure}[h]
\label{fig:2}
\includegraphics[width=\columnwidth]{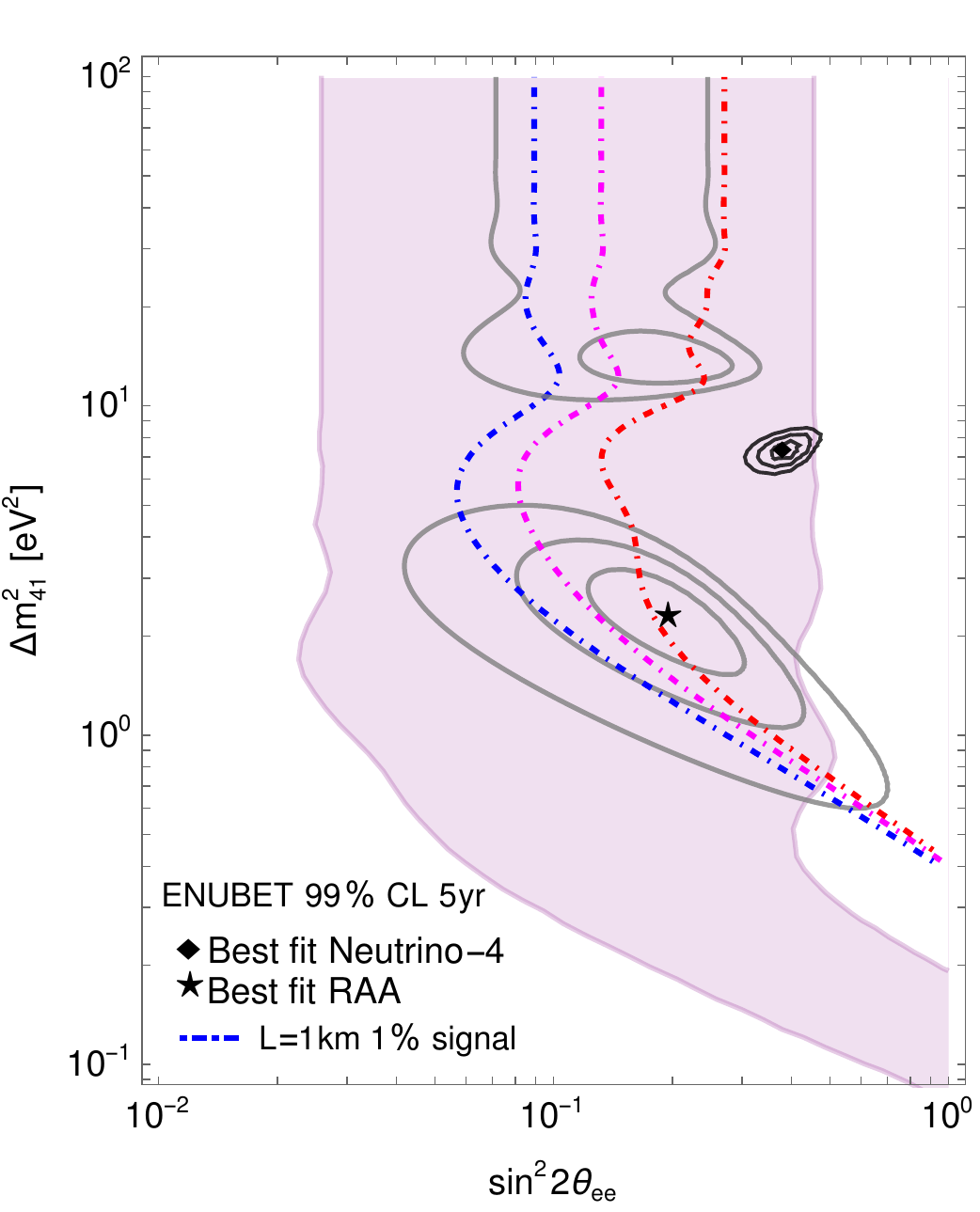}
\caption{\label{fig:2} Comparison of the expected sensitivities in the
  $\sin^2 2\theta_{ee}$-$\Delta m^2_{41}$ plane. The blue, magenta and
  red dashed-dotted lines corresponds to a 1$\%$, 2$\%$ and 5$\%$
  signal systematics at baseline length of $L=1$km. The black diamond
  point represents the best fit point from Neutrino-4
  \cite{Serebrov:2018vdw}, star represents the best fit of the reactor
  anti-neutrino anomaly RAA \cite{Mention:2011rk}. In addition we show
  90$\%$ C.L. allowed region (purple shaded area) of the gallium
  anomaly JUN45 \cite{Kostensalo:2019vmv}.}
\end{figure}

Figure [\ref{fig:2}] shows the sterile neutrino oscillation
sensitivity at ENUBET in the $\sin^2 2\theta_{ee}$-$\Delta m^2_{41}$
plane at 99$\%$ C.L. for an exposure of 1 kt assuming five years of
beam operation. The blue, magenta and red dashed-dotted lines account
for 1$\%$, 2$\%$ and 5$\%$ signal normalization systematic. Also shown
the 1$\sigma$, 2$\sigma$, 3$\sigma$ preferred regions for the best fit
RAA \cite{Mention:2011rk} and Neutrino-4 \cite{Serebrov:2018vdw}
assuming 2$\%$ signal normalization systematic.


Until recent results from IceCube \cite{PRLIce,Aartsen:2020iky} no indication
for sterile neutrino oscillation in the muon disappearance channel had
been found. In the 3$+$1 scenario with short baseline approximation,
the muon neutrino disappearance probability is given by
\begin{equation}
\label{prob3}
\begin{split}
    P_{\mu \mu} & =1-4\vert U_{\mu 4}\vert^2 \big(1- \vert U_{\mu 4}\vert^2\big) \sin^2 \big( 1.27 \Delta m^2_{41} \frac{L}{ E} \big) \\
   &= 1-\sin^2 2 \theta_{\mu \mu} \sin^2 \big( 1.27 \Delta m^2_{41} \frac{ L}{ E} \big),
\end{split}
\end{equation}
where $\vert U_{\mu 4} \vert = \cos \theta_{14}\sin \theta_{24}$ and the effective mixing angle $\theta_{ \mu \mu}$ depends on both $\theta_{14}$ and $\theta_{24}$
\begin{equation}
\begin{split}
\sin^2 2 \theta_{\mu \mu}=4 \cos^2\theta_{14}\sin^2\theta_{24}\big(1-\cos^2\theta_{14}\sin^2\theta_{24} \big).\\
\end{split}
\end{equation}

\begin{figure}[h]
\label{fig:3}
\includegraphics[width=\columnwidth]{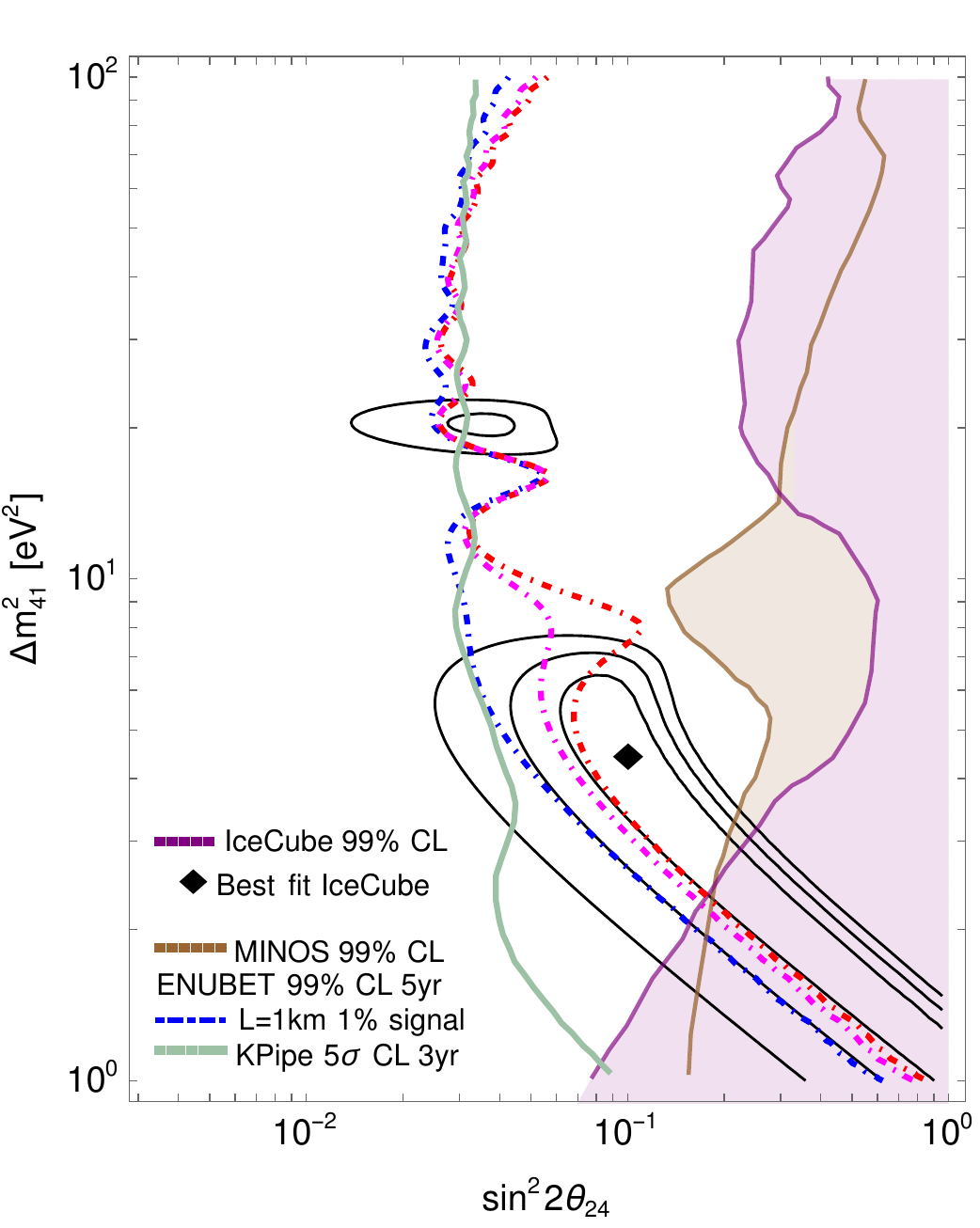}
\caption{\label{fig:3}Comparison of the expected sensitivities in the $\sin^2 2\theta_{24} $-$\Delta m^2_{41}$ plane ($\sin^2 2\theta_{\mu \mu} \approx \sin^2 2\theta_{24}$). The blue, magenta and red dashed-dotted lines corresponds to a 1$\%$, 2$\%$ and 5$\%$ signal systematics at baseline length of $L=1$km. The black diamond point represents the best fit point 90$\%$ C.L. from IceCube \cite{PRLIce, Aartsen:2020iky}, shaded purple/brown areas are excluded by IceCube and MINOS \cite{PRLMINOS} respectively.}
\end{figure}

Figure [\ref{fig:3}] shows the sterile neutrino oscillation
sensitivity at ENUBET in the $\sin^2 2\theta_{24}$-$\Delta m^2_{41}$
plane at 99$\%$ C.L. for an exposure of 1 kt assuming five years of
beam operation. The blue, magenta and red dashed-dotted lines account
for 1$\%$, 2$\%$ and 5$\%$ signal normalization systematic. Also shown
the 1$\sigma$, 2$\sigma$, 3$\sigma$ preferred regions assuming 2$\%$
signal normalization systematic for true $\sin^22 \theta_{24}=0.1$ and
$\Delta m^2_{41}=4.5$ eV$^2$ from IceCube \cite{PRLIce,Aartsen:2020iky}. For comparison we also show the sensitivity of another kaon decay based proposal, called Kpipe~\cite{Kpipe}: Kpipe  is a proposed experiment to investigate sterile neutrinos from kaon decay at rest and is aimed to set the strongest limits in the muon neutrino disappearance channel. The expected number of $\nu_{\mu}$ events is 1.02$\times10^5$ events/year in a 684 ton liquid scintillator detector, which comes in the shape of 120\,m long pipe. Kpipe would have about an order of magnitude more signal events than the  proposal we consider here and thus in principle has the potential to provide excellent sensitivity in this channel.

\section{Conclusions}\label{sec:con}
In this letter we test different hypotheses regarding simulated experimental data of the ENUBET beamline technology, we demonstrate the capabilities of tagged kaon beams in the electron and neutrino disappearance channel to investigate intriguing indications from the Neutrino-4 and IceCube collaborations. The strength of the setup considered is the vanishingly small systematic errors from the beam flux and a virtually background free measurement. The drawback is the relatively low beam luminosity, as result of the need to tag kaon decays individually. The proposed setup is envisioned as add-on measurement to the cross section program of ENUBET. The physics case arises mainly from the Neutrino-4 result, which is in a $\Delta m^2$-region which ultimately may be hard for reactor neutrino experiments to test decisively. The proposed setup could decisively test either indication, IceCube at the $5\,\sigma$ level and Neutrino-4 at the $10\,\sigma$ level. This setup is not unique in this capability and of course dedicated facilities like nuSTORM~\cite{nuSTORM} would provide superior sensitivity. In the hunt for the sterile neutrino, opportunistic measurements always have played a major role and we point out that if ENUBET is built and the Neutrino-4 indication persists, the setup in this paper would present such an opportunity.

\section{Acknowledgments}
We acknowledge useful discussions with R.~Pestes   and J. M. Berryman. This work was supported by the U.S. Department of Energy Office of Science under award number \protect{DE-SC00018327}.

\bibliography{refs,references}

\end{document}